# QUANTIFICATION OF AVOIDABLE YIELD LOSSES IN OILSEED *BRASSICA* CAUSED BY INSECT PESTS

*Jagdev Singh Kular[1]\*, Sarwan Kumar[1]\**

Punjab Agricultural University, Ludhiana, India
[1] Department of Entomology
[1] Department of Plant Breeding and Genetics

**Abstract:** A six year field study was conducted from 2001–2002 to 2006–2007 at Punjab Agricultural University, Ludhiana, India to study the losses in seed yield of different *Brassica* species (*B. juncea, B. napus, B. carinata, B. rapa* and *Eruca sativa*) by the infestation of insect pests. The experiment was conducted in two different sets viz. protected/sprayed and unprotected, in a randomized block design, with three replications. Data on the infestation of insect pests, and seed yield were recorded at weekly intervals and at harvest, respectively. The loss in seed yield, due to mustard aphid and cabbage caterpillar, varied from 6.5 to 26.4 per cent. *E. sativa* suffered the least loss in seed yield and harboured the minimum population of mustard aphid (2.1 aphids/plant) and cabbage caterpillar (2.4 larvae/plant). On the other hand, *B. carinata* was highly susceptible to the cabbage caterpillar (26.2 larvae/plant) and suffered the maximum yield loss (26.4%).

**Key words:** *Brassica* species, *Lipaphis erysimi, Pieris brassicae*, avoidable yield loss

## INTRODUCTION

India is the largest importer and third largest consumer of edible oils after China and the European Union, with a consumption of over 10 MT (Metric tones) of edible oils per year (Agarwal 2007). To make India self sufficient in oilseed production, we have to increase the production and productivity of these oilseeds. In India, these energy rich oilseed crops are grown mostly under energy deprived conditions, on marginal lands and poor soils.

Among the various oilseed crops, the oilseed *Brassica* species, also collectively called rapeseed-mustard, is composed largely of *B. juncea, B. rapa, B. napus* and *B. carinata*. Rapeseed-mustard is ranked third at the global level preceded by soybean and palm oil (Agnihotri and Prem 2007). About 44 per cent of the world's average production of rapeseed-mustard has been contributed by developing countries, in particular China and India (FAOSTAT 2007).

The yield potential of rapeseed-mustard has not been fully realized due to limits imposed upon it by a number of biotic and abiotic factors. Among the various biotic factors, the attack of insect pests is the major limiting factor in achieving higher productivity. A number of insect pests are known to attack rapeseed-mustard right from sowing till harvest. Only a few of the inects cause serious losses. According to Dhaliwal *et al.* (2004), rapeseed-mustard in India generally suffers a 30 per cent yield loss due to insect pests. This loss amounts to 27 300 million of indian rupees, annually (approximately 600 million US dollars).

The estimation of crop damage due to insect pests is critical to pest management as it helps in assigning priorities on the basis of relative importance of different pests, evaluating crop varieties for their resistance to pests and deciding the allocations for research and extension in plant protection. In Punjab, rapeseed-mustard suffers yield losses mainly due to the attack of the mustard aphid, *Lipaphis erysimi* Kaltenbach and cabbage caterpillar, *Pieris brassicae* (Linnaeus). The activity and subsequent severity of these pests varies under different agro climatic condition. For this reason,, a need was felt to generate location specific information about the amount of damage that these pests inflict on different oilseed *Brassica*.

The present study was undertaken to generate location specific information about the amount of damage inflicted on oilseed *Brassicas* by insect pests, and to identify gentotypes that suffer the least damage.

## MATERIALS AND METHODS

The study was conducted during six *rabi* crop seasons from 2001–2002 to 2006–2007 at the oilseeds research farm of the Department of Plant Breeding and Genetics, Punjab Agricultural University, Ludhiana, Punjab, India. The assessment of yield losses was done on five major groups of rapeseed-mustard viz. *Brassica juncea, B. napus, B. carinata, B. rapa* and *Eruca sativa*. The experiment was laid out in randomized block design, with two different sets viz. protected and unprotected. The crop was sown in the third week of October. Plot size was kept at 4x3 m and there were 6 replications for each *Brassica* species, with 3 each for protected and unprotected sets. The protected set was kept free from the attack of insect pests by spraying it three times with Metasystox 25 EC (oxydemeton methyl 25 EC) and Thiodan 35 EC (endosulfan 25 EC) @ 1250 ml/ha at 15 days interval. In the unprotected set, only water spray was done at the time of the insecticidal spray in the protected set.

*Corresponding address:
 js_kular@rediffmail.com, sarwanent@gmail.com

Weekly observations, for the incidence of insect pests, were made from 10 plants selected at random from each plot. At the time of harvest, data on seed yield from protected and unprotected plots were recorded, and the per cent yield losses were computed. The data on pest incidence and yield loss for the six years were pooled together to have an overall picture of the loss in yield due to pest incidence.

## RESULTS AND DISCUSSION

**Aphid population**

In 2001–2002 crop season, the population of mustard aphid varied from 0.0 to 23.8 aphids/plant in the unprotected set while the protected set was free from any infestation of insect pests. The maximum population of 23.8 aphids/plant was observed on non Canola *B. napus* variety, while there was no population on *E. sativa* (Table 1).

In 2002–2003, again the *E. sativa* was free from aphid population in the unprotected set while the maximum population of 21.1 aphids/plant was observed on *B. rapa*. It was followed by non Canola *B. napus* (20.6 aphids/plant), Canola *B. napus* (15.7 aphids/plant) and *B. carinata* (10.2 aphids/plant). In the protected set, a very low population of 0.7 and 3.9 aphids/plant was observed on Canola *B. napus* and *B. rapa*, respectively. The remaining genotypes were free from aphid infestation.

In 2003–2004, *E. sativa* harboured a minimum population of 3.3 aphids/plant in the unprotected set while the maximum population was observed on *B. juncea* (17.3 aphids/plant). It was followed by *B. rapa* (15.7 aphids/plant), non Canola *B. napus* (12.4 aphids/plant), *B. carinata* (7.5 aphids/plant) and Canola *B. napus* (7.3 aphids/plant). In the protected set a very low population of 0.7 and 1.5 aphids/plant was observed on *B. carinata* and non Canola *B. napus*, respectively. The remaining genotypes were free from aphid infestation.

In 2004–2005, *E. sativa* and *B. juncea* were not tested. Thus, among the four genotypes, a minimum population of 6.3 aphids/plant was observed on non Canola *B. napus*, while the maximum population of 32.0 aphids/plant was observed on Canola *B. napus*. It was followed by *B. rapa* (25.6 aphids/plant) and *B. carinata* (7.3 aphids/plant). In the protected set, a very negligible population of 1.5 aphids/plant was observed on Canola *B. napus*. Other genotypes were free from aphid infestation.

In 2005–2006, *E. sativa* harboured a minimum population of only 5.0 aphids/plant in the unprotected set, while the maximum population of 97.7 aphids/plant was observed on Canola *B. napus*. It was followed by *B. rapa* (92.3 aphids/plant), Canola *B. napus* (51.5 aphids/plant), *B. juncea* (38.0 aphids/plant) and *B. carinata* (20.0 aphids/plant). In the protected set, a very low population of 1.5 and 2.0 aphids/plant was observed on *B. juncea* and non Canola *B. napus*, respectively. The remaining genotypes were free from aphid infestation.

In 2006–2007, again the minimum population of 2.4 aphids/plant was observed on *E. sativa* while the maximum population was observed on Canola *B. napus* (24.9 aphids/plant). It was followed by *B. carinata* (24.6 aphids/plant), non Canola *B. napus* (20.3 aphids/plant) and *B. juncea* (12.4 aphids/plant). The protected set was free from aphid infestation.

From the pooled data of six years, it is evident that the mustard aphid population varied from 2.1 to 32.4 aphids/plant on different genotypes. The maximum population of 32.4 aphids/plant was observed on *B. rapa*. The aphid population on the genotypes *E. sativa* and *B. carinata* (2.1 and 13.7/plant, respectively) was significantly lower than that on *B. rapa*.

**Cabbage caterpillar population**

In the 2001–2002 crop season the cabbage caterpillar population varied from 2.6 to 35.8 larvae/plant in the unprotected set. The protected set was free from any cabbage caterpillar infestation. The minimum cabbage caterpillar population of 2.6 larvae/plant was observed on *E. sativa* while the maximum population on *B. rapa* was (35.8 larvae/plant) (Table 2). It was followed by non Canola *B. napus* (35.3 larvae/plant), *B. carinata* (29.8 larvae/plant), *B. juncea* (14.3 larvae/plant) and Canola *B. napus* (4.6 larvae/plant).

In 2002–2003, again the minimum population of 0.3 larvae/plant was observed on *E. sativa* while it was maximum on *B. carinata* (9.5 larvae/plant) in the unprotected set. Almost the same trend was observed in 2003–2004 and 2004–2005, however, *E. sativa* was not included in 2004–2005. In 2005–2006, *E. sativa* was free from any cabbage caterpillar population, while the maximum population was observed on *B. carinata* (29.0 larvae/plant). Almost the same trend was observed in 2006–2007, with minimum population (8.3 larvae/plant) on *E. sativa* and maximum (58.3) on *B. carinata*.

From the pooled data of six years, it is evident that the cabbage caterpillar population varied from 2.4 to 26.2 larvae/plant on different genotypes. It was maximum (26.2 larvae/plant) on *B. carinata* and minimum (2.4 larvae/plant) on *E. sativa*. The population on *B. carinata* was significantly lower than on the remaining genotypes.

**Avoidable yield loss**

In 2001–2002, the loss in the yield by insect pests varied from 3.1 to 35.3 per cent on different genotypes (Table 3). The maximum yield loss (35.3%) was observed in genotype non Canola *B. napus*. It was followed by *B. juncea* (10.7%) and *B. carinata* (9.2%). The remaining three genotypes suffered much less yield loss. Canola *B. napus*, *E. sativa* and *B. rapa*, had a yield loss of 5.7, 5.2 and 3.2 per cent, respectively. Tables 1, 2 show the aphid and cabbage caterpillar population numbers. From the data in these tables it is evident that the 35.3 per cent yield loss in non Canola *B. napus* corresponded with the high aphid population of 23.8 aphids/plant, and high cabbage caterpillar population of 35.3 larvae/plant under unsprayed condi-

Table 1.  Number of *L. erysimi* on different *Brassica* species

| *Brassica* spp. | Genotype | 2001–2002 | | 2002–2003 | | 2003–2004 | | 2004–2005 | | 2005–2006 | | 2006–2007 | | mean | |
|---|---|---|---|---|---|---|---|---|---|---|---|---|---|---|---|
| | | P* | UP** | P | UP | P | UP | P | UP | P | UP | P | UP | P | UP |
| *B. juncea* | PBR-91 | 0.0 | 2.6 | – | – | 0.0 | 17.3 | – | – | 1.5 | 38.0 | 0.0 | 12.4 | 0.4 | 17.6 |
| *B. napus* | GSL-1 (non Canola) | 0.0 | 23.8 | 0.0 | 20.6 | 1.5 | 12.4 | 0.0 | 6.3 | 2.0 | 51.5 | 0.0 | 20.3 | 0.6 | 22.5 |
| | GSC-5 (Canola) | 0.0 | 9.2 | 0.7 | 15.7 | 0.0 | 7.3 | 1.5 | 32.0 | 0.0 | 97.7 | 0.0 | 24.9 | 0.4 | 31.1 |
| B. carinata | PC-5 | 0.0 | 12.5 | 0.0 | 10.2 | 0.7 | 7.5 | 0.0 | 7.3 | 0.0 | 20.0 | 0.0 | 24.6 | 0.1 | 13.7 |
| *B. rapa* var. brown sarson | BSH-1 | 0.0 | 7.4 | 3.9 | 21.1 | 0.0 | 15.7 | 0.0 | 25.6 | 0.0 | 92.3 | – | – | 0.8 | 32.4 |
| *E. sativa* | TMLC 2 | 0.0 | 0.0 | 0.0 | 0.0 | 0.0 | 3.3 | – | – | 0.0 | 5.0 | 0.0 | 2.4 | 0.0 | 2.1 |

* P – protected; **UP – unprotected

Table 2.  Number of *P. brassicae* larvae on different *Brassica* species

| *Brassica* spp. | Genotype | 2001–2002 | | 2002–2003 | | 2003–2004 | | 2004–2005 | | 2005–2006 | | 2006–2007 | | mean | |
|---|---|---|---|---|---|---|---|---|---|---|---|---|---|---|---|
| | | P* | UP** | P | UP | P | UP | P | UP | P | UP | P | UP | P | UP |
| *B. juncea* | PBR-91 | 0.0 | 14.3 | – | – | 0.7 | 4.5 | – | – | 0.0 | 7.0 | 0.0 | 21.6 | 0.2 | 11.9 |
| *B. napus* | GSL-1 (non Canola) | 0.0 | 35.3 | 1.0 | 7.7 | 0.0 | 1.3 | 0.0 | 1.5 | 0.0 | 10.1 | 0.0 | 33.3 | 0.2 | 14.9 |
| | GSC-5 (Canola) | 0.0 | 4.6 | 0.0 | 6.3 | 0.0 | 3.0 | 0.0 | 2.7 | 0.0 | 1.7 | 0.0 | 28.3 | 0.0 | 7.8 |
| B. carinata | PC-5 | 0.0 | 29.8 | 0.5 | 9.5 | 2.0 | 13.3 | 0.0 | 17.3 | 0.0 | 29.0 | 0.0 | 58.3 | 0.4 | 26.2 |
| *B. rapa* var. brown sarson | BSH-1 | 0.0 | 35.8 | 0.0 | 3.7 | 0.9 | 3.7 | 0.0 | 5.3 | 0.0 | 20.0 | – | – | 0.2 | 13.7 |
| *E. sativa* | TMLC 2 | 0.0 | 2.6 | 0.0 | 0.3 | 0.0 | 0.7 | – | – | 0.0 | 0.0 | 0.0 | 8.3 | 0.0 | 2.4 |

* P – protected; ** UP – unprotected

Table 3. Loss in seed yield due to damage by insect pests, in different *Brassica* species

| *Brassica* spp. | Genotype | Per cent loss in seed yield | | | | | | |
|---|---|---|---|---|---|---|---|---|
| | | 2001–2002 | 2002–2003 | 2003–2004 | 2004–2005 | 2005–2006 | 2006–2007 | mean |
| *B. juncea* | PBR-91 | 10.7 | – | 5.7 | – | 4.8 | 5.2 | 6.8 |
| *B. napus* | GSL-1 (non Canola) | 35.3 | 32.6 | 1.5 | 4.6 | 18.0 | 19.7 | 20.7 |
| | GSC-5 (Canola) | 5.7 | 21.5 | 10.0 | 22.5 | 17.6 | 8.2 | 11.6 |
| *B. carinata* | PC-5 | 9.2 | 24.8 | 16.7 | 57.3 | 33.2 | 19.8 | 26.4 |
| *B. rapa* var. brown sarson | BSH-1 | 3.1 | 33.4 | 9.7 | 28.8 | 15.9 | – | 16.7 |
| *E. sativa* | TMLC 2 | 5.2 | 0.0 | 2.9 | – | 13.3 | 4.8 | 6.5 |

Table 4. Population of insect pests and seed yield in different *Brassica* species (pooled data from over a period of 6 years)

| *Brassica* spp. | Genotype | Aphid population/plant | | Cabbage caterpillars/plant | | Yield [q/ha] | | Yield loss [%] |
|---|---|---|---|---|---|---|---|---|
| | | P* | UP** | P | UP | P | UP | |
| *Brassica juncea* | PBR-91 | 0.4 | 17.6 | 0.2 | 11.9 | 20.3 | 18.9 | 6.8 |
| *B. napus* | GSL-1 (non Canola) | 0.6 | 22.5 | 0.2 | 14.9 | 14.9 | 11.8 | 20.7 |
| | GSC-5 (Canola) | 0.4 | 31.1 | 0.0 | 7.8 | 16.8 | 14.4 | 11.6 |
| *B. carinata* | PC-5 | 0.1 | 13.7 | 0.4 | 26.2 | 16.7 | 12.3 | 26.4 |
| *B. rapa* var. brown sarson | BSH-1 | 0.8 | 32.4 | 0.2 | 13.7 | 9.7 | 8.1 | 16.7 |
| *E. sativa* | TMLC 2 | 0.0 | 2.1 | 0.0 | 2.4 | 5.5 | 5.2 | 6.5 |
| CD (p = 0.05) | | NS | 18.6 | NS | 11.3 | 5.04 | 4.7 | |

* P – protected; ** UP – unprotected

tions, in 2001–2002. Similarly, the low yield loss (5.2%) in *E. sativa* corresponded with no aphid population and the low cabbage caterpillar population (2.6 larvae/plant). The yield loss in *B. rapa* was minimum (3.1%), although it harboured 35.8 larvae/plant of cabbage caterpillar.

In 2002–2003, the yield loss in different genotypes varied from 0.0 to 33.4 per cent. The yield loss was maximum in *B. rapa* followed by non Canola *B. napus* (32.6%), *B. carinata* (24.8%) and Canola *B. napus* (21.5%). However, no yield loss was observed in *E. sativa*. In genotype *B. rapa*, the maximum yield loss corresponded with the maximum population of mustard aphid (21.1 aphids/plant) and 3.7 larvae/plant of cabbage caterpillar under unsprayed conditions. No population of mustard aphid was observed in *E. sativa*, while the population of the cabbage caterpillar was only 0.3 larvae/plant. Consequently, there was no yield loss in this genotype.

The yield loss in 2003–2004 varied from 1.5 to 16.7 per cent on different genotypes. The genotype *B. carinata* recorded the maximum yield loss (16.7%). It was followed by Canola *B. napus* (10.0%), *B. rapa* (9.7%), *B. juncea* (5.7%), *E. sativa* (2.9%) and non Canola *B. napus* (1.5%). In *B. carinata*, the maximum yield loss corresponded with the maximum population of cabbage caterpillar (13.3 larvae/plant) under unsprayed conditions, while the population of mustard aphid was 7.5 aphids/plant. Similarly, the only 2.9 per cent yield loss in *E. sativa* corresponded with a minimum population of cabbage caterpillar (0.7 larvae/plant) and low population of mustard aphid (3.3 aphids/plant).

In 2004–2005, the loss in seed yield varied from 4.6 per cent to as high as 57.3 per cent. The maximum loss (57.3%) in *B. carinata* corresponded with the maximum population of cabbage caterpillar (17.3 larvae/plant) under unsprayed conditions, while the population of mustard aphid was 7.3 aphids/plant. The high loss in yield *i.e.* 28.8 and 22.5 per cent in *B. rapa* and Canola *B. napus* corresponded with high aphid population i.e. 25.6 and 32.0 aphids/plant, respectively. The loss in yield was minimum (4.6%) in non Canola *B. napus*, which corresponded with the minimum population of mustard aphid (6.3 aphids/plant) as well as cabbage caterpillar (1.5 larvae/plant).

In 2005–2006, the yield loss in different species varied from 4.8 to 33.2 per cent. Similar to what had been observed in the previous year, the maximum loss in seed yield in *B. carinata* (33.2%) corresponded with the maximum population of cabbage caterpillar (29.0 larvae/plant) while the population of mustard aphid was 20.0 aphids/plant. It was followed by non Canola *B. napus*, Canola *B. napus* and *B. rapa* where the respective loss in seed yield was 17.9, 17.6 and 15.9 per cent. The yield loss in these genotypes corresponded with high aphid population *i.e.* 51.5, 97.7 and 92.3 aphids/plant in non Canola *B. napus*, Canola *B. napus* and *B. rapa*, respectively. The respective population of cabbage caterpillar was 10.7, 1.7 and 20.0 larvae/plant in these genotypes.

In 2006–2007, the loss in yield in various genotypes varied from 4.8 to 19.8. The genotype *B. carinata* suffered the maximum loss in seed yield (19.8%) which corresponded with the maximum population of cabbage caterpillar (58.3 larvae/plant) and the high population of mustard aphid (24.6 aphids/plant). It was followed by non Canola *B. napus* and Canola *B. napus* (19.7 and 8.2% yield loss, respectively). The respective population of mustard aphid and cabbage caterpillar in these genotypes was 20.3 and 24.9 aphids/plant and 33.3 and 28.3 larvae/plant. The minimum loss in seed yield (4.8%) in *E. sativa* corresponded with the minimum population of mustard aphid (2.4 aphids/plant) and cabbage caterpillar (8.3 larvae/plant).

**Pooled data**

Six years of pooled data on the loss in seed yield and population of the mustard aphid and cabbage caterpillar on different genotypes are presented in table 4. On the basis of six years of experiments, it can be stated that losses in various genotypes due to insect pests varied from 6.5 to 26.4 per cent. The genotype *B. carinata* suffered the maximum loss in seed yield (26.4%). It was followed by non Canola *B. napus* (20.7% loss), *B. rapa* (16.7%) and Canola *B. napus* (11.6%). The genotypes *B. juncea* and *E. sativa* suffered less than a 7 per cent loss in seed yield.

The population of the mustard aphid varied from 2.1 to 32.4 aphids/plant under unprotected conditions. This aphid population was significantly lower in *E. sativa* than that in non Canola *B. napus*, Canola *B. napus* and *B. rapa*. The population of the cabbage caterpillar varied from 2.4 to 26.2 larvae/plant. The minimum population was observed in *E. sativa* (2.4 larvae/plant), which was significantly lower than non Canola *B. napus*, *B. carinata* and *B. rapa*. The maximum population of cabbage caterpillar (26.2 larvae/plant) was recorded for *B. carinata*.

From the data it was evident that maximum yield loss (26.4%) in *B. carinata* corresponded with the maximum population of cabbage caterpillar (26.2 larvae/plant), while the population of mustard aphid was 13.7 aphids/plant. On the other hand, the minimum yield loss (6.5%) in *E. sativa* corresponded with the minimum population of mustard aphid (2.1 aphids/plant) as well as cabbage caterpillar (2.4 larvae/plant).

Patel *et al.* (2004) reported the losses in seed yield of Indian mustard, *B. juncea* by the mustard aphid to be 76.0 to 100.0 per cent under unsprayed conditions. Singh and Sachan (1994) also reported avoidable losses due to mustard aphid which were up to 69.6 per cent. Similarly, Bakhetia (1984) observed a 57.8 to 80.6 per cent yield loss due to mustard aphid. In Haryana, Singh *et al.* (1993) reported 38.20 to 46.56 per cent yield losses in susceptible cultivars under unprotected conditions. However, no information was available in literature about the yield losses in oilseed *Brassicas*, by cabbage caterpillar.

## CONCLUSION

Thus, from the present study it can be concluded that yield losses in oilseed *Brassica* by insect pests, particularly mustard aphid and cabbage caterpillar, vary from 6.5 to 26.4 per cent. *E. sativa* harboured a significantly lower population of mustard aphid and cabbage caterpillar than the remaining genotypes. *B. carinata* was highly susceptible to the attack of cabbage caterpillar and *B. rapa*, Canola *B. napus* and non Canola *B. napus* were susceptible to the attack of cabbage caterpillar.

## POLISH SUMMARY

### OCENA STRAT PLONU NASION ROŚLIN OLEISTYCH Z RODZAJU *BRASSICA* POWODOWANYCH PRZEZ SZKODNIKI

W latach 2001–2007 na Uniwersytecie Rolniczym Penjab w Ludhianie (Indie), prowadzono 6-letnie badania polowe nad określeniem strat w plonach nasion różnych gatunków roślin, należących do rodzaju *Brassica* (*Brassica junicea*, *B. carinata*, *B. rapa*, *Eruca sativa*), spowodowanych przez szkodniki. Doświadczenie polowe założono metodą bloków losowanych w trzech powtórzeniach, z uwzględnieniem dwóch wariantów – rośliny chronione/opryskiwane oraz kontrola – rośliny nie chronione. Obserwacje nasilenia występowania szkodliwych owadów prowadzono z tygodniowymi przerwami, a dane dotyczące plonu uzyskano po zbiorze nasion. Straty plonu nasion spowodowane opanowaniem roślin przez mszycę *Lipaphis erysimi* oraz gąsienice bielinka kapustnika *Pieris brassicae* wynosiły 6,6–26,4%. Najmniejsze straty w plonach nasion stwierdzono w przypadku roślin rukoli (*Eruca sativa*), opanowanych w niewielkim stopniu przez populacje mszycy *L. erysimi* (2,1 mszyc/roślinę) i gąsienice kapustnych (2,4 larw/roślinę). Rośliny *B. carinata* były bardzo podatne na żerowanie gąsienic kapustnych (26.2 larw/roślinę) i reagowały maksymalną stratą plonu (26,4%).